\documentclass[amsmath,superscriptaddress,aps,twocolumn,showpacs,prl]{revtex4-1}

\usepackage{bm}
\usepackage{graphics}
\usepackage{graphicx}
\usepackage{amsthm}
\usepackage{amsmath}
\usepackage{wasysym}
\usepackage{amssymb}
\usepackage[dvips]{epsfig}

\begin{document}
\title{Exact Chiral Spin Liquid with Stable Spin Fermi Surface on the Kagome Lattice}
\author{Victor Chua}
\affiliation{Department of Physics, The University of Texas at Austin, Austin, TX 78712, USA}
\author{Hong Yao}
\affiliation{Department of Physics, University of California, Berkeley, California 94720, USA}
\affiliation{Materials Sciences Division, Lawrence Berkeley National Laboratory, Berkeley, California 94720, USA}
\author{Gregory A. Fiete}
\affiliation{Department of Physics, The University of Texas at Austin, Austin, TX 78712, USA}
\begin{abstract}
We study an exactly solvable quantum spin model of Kitaev type on the kagome lattice.  We find a rich phase diagram which includes a topological (gapped) chiral spin liquid with 
gapless chiral edge states, and a gapless chiral spin liquid phase with a spin Fermi surface. The ground state of the current model contains an odd number of electrons per unit cell which qualitatively distinguishes it from previously studied exactly solvable models with a spin Fermi surface. Moreover, we show that the spin Fermi surface is {\it stable} against weak perturbations. 
\end{abstract}



\maketitle

\global\long\def\Ju{J_{\triangle}}
\global\long\def\Jd{J_{\nabla}}
\global\long\def\ij{\langle ij\rangle}

{\it Introduction--}The search for exotic ground states of quantum many-body systems has been at the fore in experimental and theoretical research for a number of decades, 
yet nature continues to tantalize us with ever more interesting behaviors. In recent years, 
systems with some type of topological order beyond Landau's broken symmetry principle \cite{kivelson1987,wen1989}, such as the fractional quantum Hall effect \cite{laughlin1983} and quantum spin liquids \cite{anderson1973},  have garnered a great deal of attention. 
A quantum spin liquid (QSL) is an insulating state that does not exhibit any conventional magnetic order at zero temperature, typically due to a delicate balance of competing and/or ``frustrating'' effects (for a review, see  Ref. \cite{Lee:sci08,Balents:nat10}).

Various flavors of QSLs with different types of quasiparticle excitations and braiding statistics have been proposed \cite{anderson1987,kalmeyer1987,Lee:rmp06}. However, the reliability of the approximations typically invoked in their solution are often imperfectly understood. 
Thus, it is useful to find exactly solvable (even though often contrived) models with spin liquid ground states so as to establish their stability as phases of matter and to understand their essential physical properties.  The existence of such states is compellingly established \cite{Moessner:prl01} for quantum dimer models \cite{Rokhsar:prl88} on the triangular lattice.

More recently, the discovery by Kitaev \cite{Kitaev:ap06} of an exactly solvable, interacting two-dimensional spin-1/2 model on the honeycomb lattice with a spin liquid phase came as a great boon to the study of exotic ground states in magnetic systems. Since then, variants \cite{Mandal:prb09} of Kitaev's model on trivalent lattices have appeared which realize chiral spin liquids with non-Abelian anyons \cite{Yao:prl07} and spin liquids with a spin Fermi surface \cite{Yao:prl09,Baskaran:09,Tikhonov:prl10}. They have also been used  to study quantum critical points \cite{Feng:prl07,Chung:prb10}, entanglement entropy and entanglement spectrum \cite{Yao:prl10}, and edge solitons \cite{Lee:prl07} among others \cite{Chen:prb07,Vidal:prb08}. If the coordination number of the lattice is larger than three, a generalization of the Kitaev model, the so-called $\Gamma$-matrix model \cite{Wen:prd03,Yao:prl09,Wu:prb09,Ryu:prb09,Nussinov:prb09,Chern:prb10}, with extra degrees of freedom on each site is needed to ensure exact solvability.   These additional degrees of freedom can be either interpreted as an ``orbital" degree of freedom \cite{Wang:prb06,Ryu:prb09}, or one may consider each site as having a larger spin, but no orbital degrees of freedom \cite{Murakami:prb04,Yao:prl09}.

\begin{figure}[b]
\includegraphics[width=7cm]{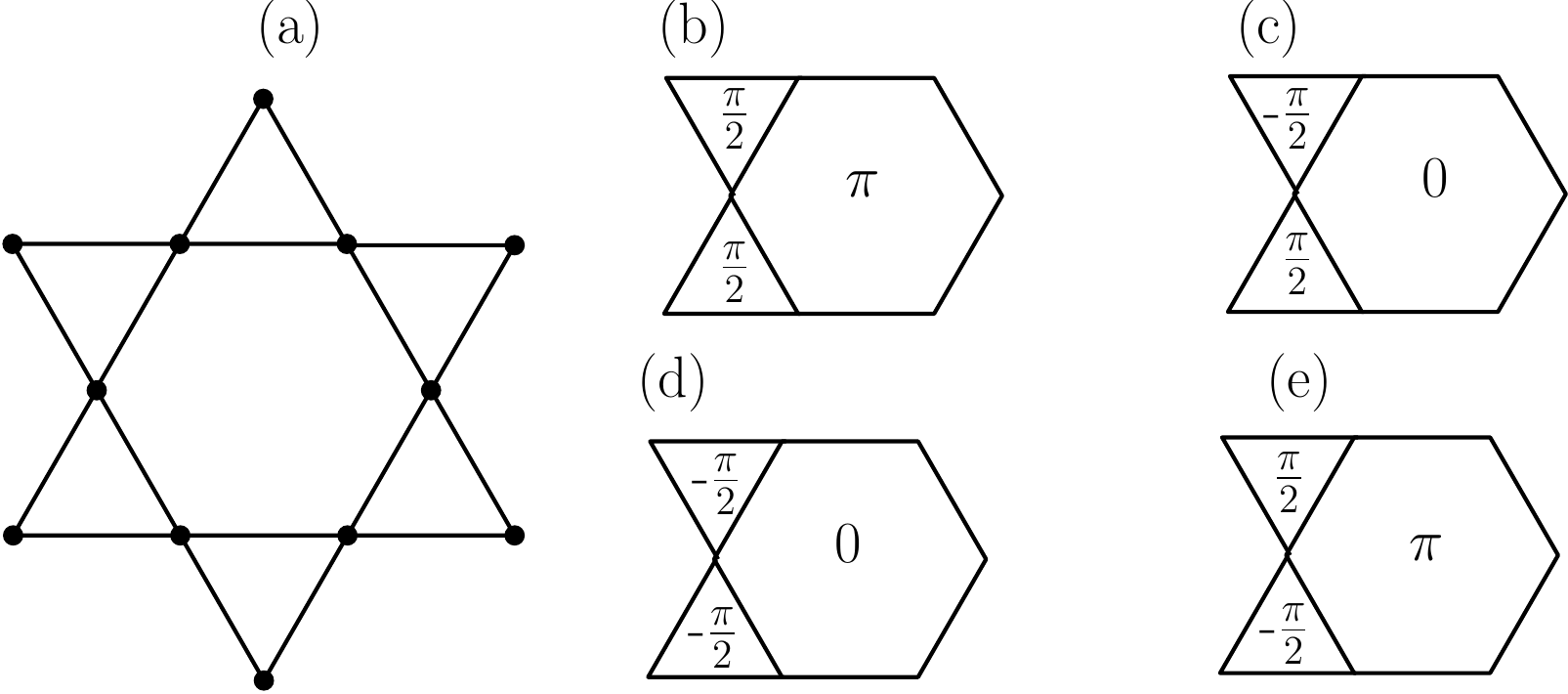}
\caption{(a) is a schematic representation of kagome lattice which is a network of corner sharing triangles in two-dimensions with 3 sites per unit cell.  (b-e) are 4 energetically distinct flux configurations preserving the translational symmetry of the lattice. A doubled (magnetic) unit cell is required for (d) and (e) where the total flux through a unit cell is $\pm\pi$. }\label{fig1}
\end{figure}

Partly motivated by the recent interest in finding a spin liquid with a spinon Fermi surface in Herbertsmithite--a Mott insulator on the kagome lattice \cite{Helton:prl07}--we study an exactly solvable spin-3/2 model on the kagome lattice (see Fig. \ref{fig1}).  Our main finding is that our model possesses a gapless spin liquid phase with a finite spin Fermi surface. We show that the spin Fermi surface is {\it stable} against any weak perturbations. The stability follows from the lack of inversion symmetry, $\pi$ rotation and time-reversal symmetry (that is spontaneously broken)
in the ground state. Both the {\it odd} number of electrons per unit cell in the ground state and the stability of spin Fermi surface qualitatively distinguishes our model from other known exactly solvable models with a finite spin Fermi surface \cite{Yao:prl09,Baskaran:09,Tikhonov:prl10}.  We also find a gapped Abelian chiral spin liquid phase \cite{Yao:prl07}, which possesses two gapless chiral Majorana edge states for a lattice geometry with boundary.

{\it Model--} The $\Gamma$-matrix model describing  $S=3/2$ spins \cite{Yao:prl09} we study respects the translational and threefold rotational symmetry of the kagome lattice and is given by
\begin{eqnarray}
\label{eq:H_gen}
{\cal H}  &=&  \Ju\sum_{\ij\in\triangle}\Gamma_{i}^{1}\Gamma_{j}^{2}+\Jd\sum_{\ij\in\nabla}\Gamma_{i}^{3}\Gamma_{j}^{4}\nonumber +J_{5}\sum_{i}\Gamma_{i}^{5}\\
&&+\Ju'\sum_{\ij\in\triangle}\Gamma_{i}^{15} \Gamma_{j}^{25} +\Jd'\sum_{\ij\in\nabla}\Gamma_{i}^{35}\Gamma_{j}^{45},
 \end{eqnarray}
where we have distinguished the nearest neighbor couplings $J_{ij}$ as $\Ju$,
$\Jd$ and $J'_{ij}$ as $\Ju'$ and $\Jd'$ based on whether the link $\ij$ belongs to an up ($\triangle$) triangle or down ($\nabla$) triangle, and $\ij$ is taken in the counterclockwise
sense for each triangle.  Locally, the five $\Gamma$-matricies satisfy a Clifford algebra, $\{\Gamma_i^a,\Gamma_i^b\}=2\delta^{ab}$, where $a,b=1,...,5$, 
and $\Gamma^{ab}\equiv [\Gamma^a,\Gamma^b]/(2i)$.  In terms of the components of the spin $S=3/2$ operators \cite{Murakami:prb04,Yao:prl09},
\begin{eqnarray}
\label{eq:Gamma_S}
\Gamma^1=\frac{1}{\sqrt 3}\{S^y,S^z\}, \Gamma^2=\frac{1}{\sqrt 3}\{S^z,S^x\}, 
\Gamma^3=\frac{1}{\sqrt 3}\{S^x,S^y\},\nonumber\\
\Gamma^4=\frac{1}{\sqrt 3}[(S^x)^2-(S^y)^2],~ 
\Gamma^5=(S^z)^2-\frac{5}{4}.\qquad~~
\end{eqnarray}
With the identification \eqref{eq:Gamma_S}, it is clear the model \eqref{eq:H_gen} has a global Ising spin symmetry under 180$^\circ$ rotations about the $z$-axis, and possesses time-reversal symmetry (TRS), although TRS will be spontaneously broken in the ground state as we describe below, in addition to the translational and threefold rotational lattice symmetry mentioned above.

{\it Method of Solution--}The key to the exact solvability \cite{Kitaev:ap06} of the model \eqref{eq:H_gen} is an infinite number of conserved operators, $\hat W_p$, that satisfy $[\hat W_p,{\cal H}]=0$ and $[\hat W_p,\hat W_{p'}]=0$ for all elementary plaquettes labeled by $p$ and $p'$. As shown in Fig. \ref{fig1}, the kagome lattice contains elementary up and down triangular plaquettes, and hexagonal ones.  Specifically, $\hat W_\triangle=\Gamma_i^{12} \Gamma_j^{12}\Gamma_k^{12}$, $\hat W_\nabla=\Gamma_i^{34} \Gamma_j^{34}\Gamma_k^{34}$, and $\hat W_{\hexagon}=\Gamma_i^{23}\Gamma_j^{14}\Gamma_k^{23}\Gamma_l^{14}\Gamma_m^{23}\Gamma_n^{14}$, where the sites are taken in a counter-clockwise fashion and for $\hat W_{\hexagon}$ the first link $(ij)$ is assumed to lie on a $\triangle$. 

In order to solve the Hamiltonian \eqref{eq:H_gen} we introduce a Majorana representation of the $\Gamma$-matricies \cite{Yao:prl07},
 \begin{equation}
 \label{eq:Gamma_Majorana}
\Gamma_{i}^{a}  =  i\xi_{i}^{a}c_{i},\quad
\Gamma_{i}^{5}  =  ic_{i}d_{i},\quad
\Gamma_{i}^{a5}  =  i\xi_{i}^{a}d_{i},
\end{equation}
with $a=1,2,3,4.$  There are thus 6 Majorana species on each site $i$: $\{\xi_{i}^{1},\xi_{i}^{2},\xi_{i}^{3},\xi_{i}^{4},c_{i},d_{i}\}$.  The Majorana representation enlarges the spin-3/2 Hilbert space, so that one must enforce the constraint $D_{i}=-\Gamma_i^1 \Gamma_i^2\Gamma_i^3\Gamma_i^4\Gamma_i^5 =-i\xi_{i}^{1}\xi_{i}^{2}\xi_{i}^{3} \xi_{i}^{4}c_{i}d_{i}=1$, namely for any physical state $|\Psi\rangle_\textrm{phys}$ $D_i|\Psi\rangle_\textrm{phys}= |\Psi\rangle_\textrm{phys}$ for any $i$.  From the enlarged Hilbert space, physical states are obtained by applying the projection operator: $P=\prod_{i}\left[\frac{1+D_{i}}{2}\right]$, where the product is over all sites in the lattice.  

Using the relations \eqref{eq:Gamma_Majorana}, the Hamiltonian \eqref{eq:H_gen} becomes
\begin{eqnarray}
\label{eq:H_Majorana_gen}
{\cal\tilde{H}}  =  \sum_{\ij}\left[J_{ij}iu_{ij}c_{i}c_{j} +J'_{ij}iu_{ij}d_{i}d_{j}\right] +J_{5}\sum_{i}ic_{i}d_{i},
 \end{eqnarray}
 where $u_{ij}=-u_{ji}$, with $u_{ij}=-i\xi_{i}^{1}\xi_{j}^{2}  \;\text{if}\;ij\in\triangle$ and $u_{ij}=
 -i\xi_{i}^{3}\xi_{j}^{4}  \;\text{\text{if}\;\ensuremath{ij\in\nabla}}$.  The original Hamiltonian is obtained  through ${\cal H}=P {\cal \tilde H}P$.   Since $[u_{ij},{\cal \tilde H}]=[u_{ij},u_{i'j'}]=0$, the original spin Hamiltonian has been reduced to a model of free Majorana fermions moving in a {\em static} background of $\mathbb{Z}_{2}$ gauge fields \cite{Kitaev:ap06}:  in \eqref{eq:H_Majorana_gen} the $u_{ij}$ can be replaced by their eigenvalues $\pm 1$. It remains to determine the pattern of the $u_{ij}$ (up to gauge transformations) that yields the lowest energy state.  We will limit our discussion to positive $J$ couplings only.

In terms of static $Z_2$ gauge fields, we define a flux $\phi_p$ via $\exp(i\phi_p)\equiv \prod_{jk\in p} iu_{jk}$, where $jk$ is taken counterclockwise, on each elementary plaquette $p$. It is clear that $\phi_p=\pm \pi/2$ in the triangular plaquettes, $\phi_p=0,\pi$ in the hexagonal plaquettes. Since  $W_{p}=\prod_{\ij\in p}u_{ij}$, where $ij$ is also taken counterclockwise, $\exp(i\phi_p)=-W_p$ ($-iW_p$) when $p$ is hexagon (triangular) plaquettes. Under time reversal symmetry, $W_p\to \pm W_p$ where $+(-)$ is for hexagon (triangle) plaquettes; it follows that $\phi_p \to -\phi_p$ for triangle plaquettes while $\phi_p$ remains unchanged for hexagon plaquettes. Consequently, a ground state with a certain flux pattern $\{\phi_p\}$ (regardless of any particular choice of $u_{ij}$) spontaneously breaks time reversal symmetry and its energy must be degenerate with the ground state with flux pattern obtained from $\{\phi_p\}$ by changing $\phi_p\to-\phi_p$ on all triangular plaquettes \cite{Yao:prl07}. 

\begin{figure}[bt]
\includegraphics[width=9cm]{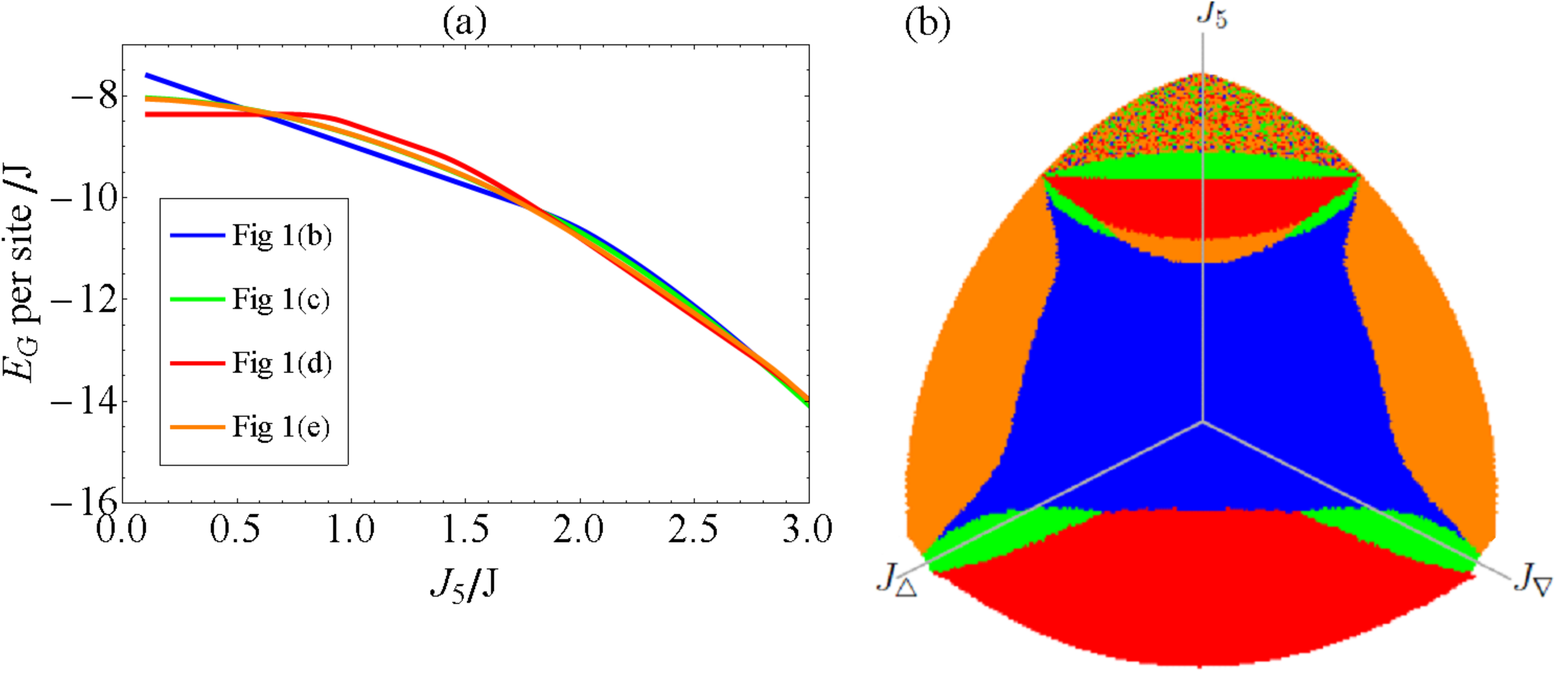}
\caption{
(color online) (a) Ground-state energy per site 
of \eqref{eq:H_gen} as a function of $J_5/J$ with $\Ju=\Ju'=\Jd=\Jd'=J$ for the uniform flux configurations shown in Fig.\ref{fig1}(b-e). (b) Ground-state flux configuration among Fig.\ref{fig1}(b-e) as a function of $\Ju,\Jd,J_5$ for $\Ju^2+\Jd^2+J_5^2=1$ with $\Jd=\Jd'$ and $\Ju=\Ju'$. The coloring scheme follows that of the legend in Fig.\ref{fig4}(a). 
} \label{fig4}
\end{figure}

In general, the nature of a ground state and its excitations crucially depends on the flux pattern $\{\phi_p\}$ in the ground state. As we will see below, the richness of this model is due in part to the freedom to add additional terms to \eqref{eq:H_gen} that preserve the exact solvability, but can be used to select different (gauge-inequivalent) configurations of $u_{ij}$ as the ground state. 

\begin{figure}[bt]
\includegraphics[width=5.5cm]{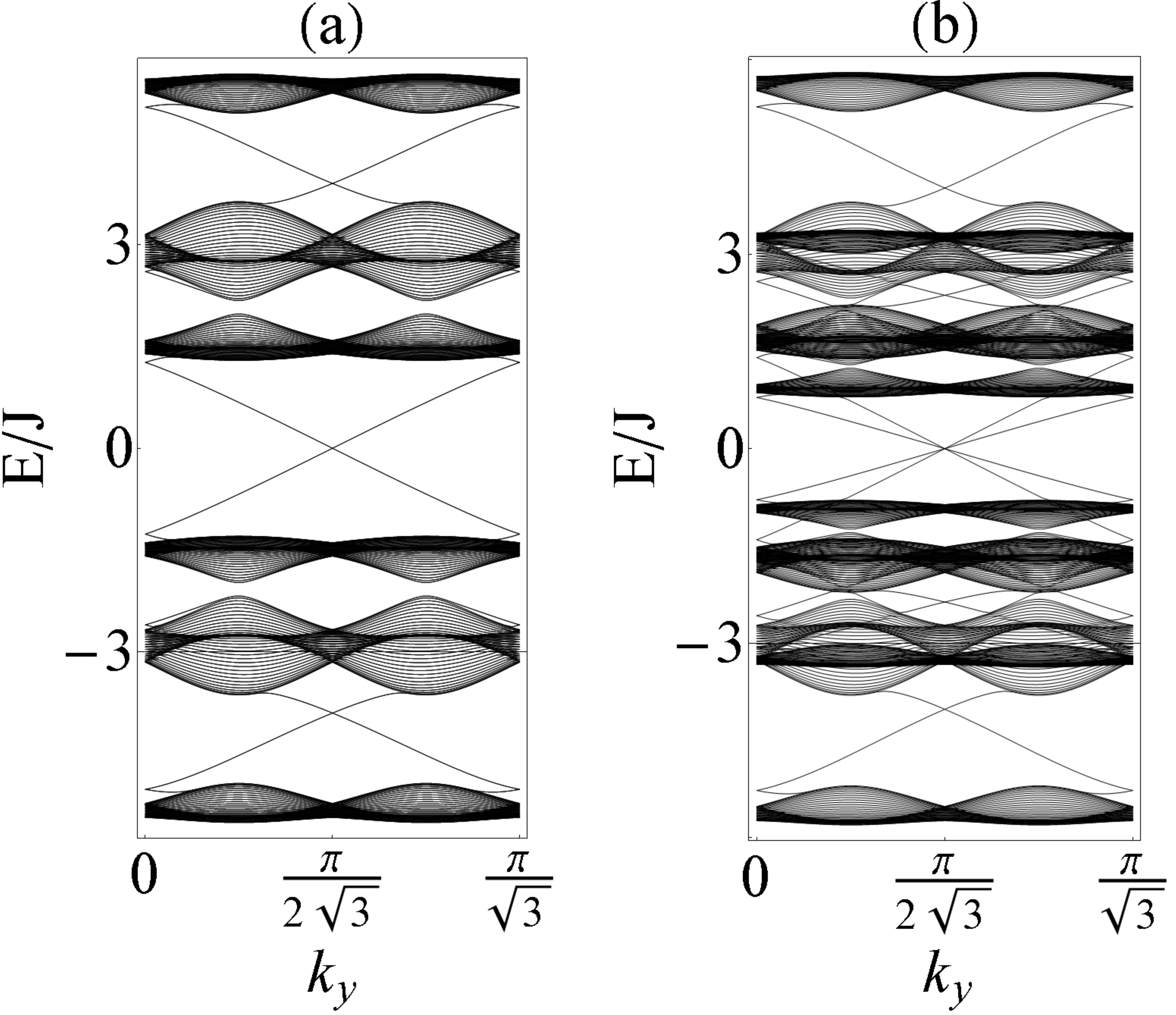}
\caption{(a) Band structure on a cylindrical geometry for $\Ju=\Ju'=1.0$, $\Jd=\Jd'=0.8$, $J_5=0$. There are two gapless chiral Majorana edge states (dotted lines) which overlap on each other. (b) For $\Ju=1.0$, $\Ju'=0.6$, $\Jd=0.9$, $\Jd'=0.5$, and $J_5=0.1$, the two gapless edges states separate. These ground states are thus CSLs with a spectrum Chern number ($\pm 2$) and the vortices obey Abelian statistics.} \label{edge}
\end{figure}

{\it Chiral Spin Liquid--}A chiral spin liquid (CSL) breaks time-reversal symmetry spontaneously and has no conventional magnetic order \cite{kalmeyer1987, Yao:prl07, Schroeter:prl07}.  As we have discussed, the current model \eqref{eq:H_gen} indeed breaks TRS spontaneously.   It remains to be shown that a ground state exists without breaking spin-rotational or translational symmetry.  To determine the ground-state flux configuration, we compute the ground-state energy $E_G$ of the Hamiltonian \eqref{eq:H_Majorana_gen} as a function of the $u_{ij}$. 

For a bipartite lattice, the ground state flux pattern is determined by the Lieb theorem \cite{Lieb:prl94} and the flux is always uniform.  However, the kagome lattice is not bipartite, so we must determine it numerically and analytically in some special limits \cite{Yao:prl07}. By diagonalizing systems of up to about $10^4$ sites and computing $E_G$, we indeed find that for various possible parameters in Hamiltonian (1), the ground state has a uniform flux pattern. There are four possible time-reversal-inequivalent uniform flux configuration labeled by $\{\phi_{\triangle},\phi_{\nabla},\phi_{\hexagon}\} =\{\frac{\pi}{2},\frac{\pi}{2},\pi\}, \{-\frac{\pi}{2},\frac{\pi}{2},\pi\}, \{\frac{\pi}{2},\frac{\pi}{2},0\}, \{-\frac{\pi}{2},\frac{\pi}{2},0\}$ as shown in Fig. 1(b-e). The ground state flux pattern depends on the relative size of all the coupling constants in Eq.\eqref{eq:H_gen}, as seen in Fig.~\ref{fig4}. For instance, when $J_5\ll \{\Ju,\Jd,\Ju',\Jd'\}$, a uniform flux state of $\{\frac{\pi}{2},\frac{\pi}{2},0\}$ or $\{-\frac{\pi}{2},-\frac{\pi}{2},0\}$ is favored.

For small $J_5$, the ground state has the uniform flux configuration $\{\frac{\pi}{2},\frac{\pi}{2},0\}$ shown in Fig.\ref{fig1}(d) and all its excitations are fully gapped for a system with periodic boundary conditions. The total flux per lattice unit cell is $\pi$; it follows that the magnetic unit cell must be doubled by a gauge choice.  Note that even though a specific gauge choice $u_{ij}$ with this flux configuration would ``seemingly'' double the unit cell and break the threefold rotational symmetry of the lattice, the physical ground state obtained after the projection is translationally and threefold rotationally invariant because the gauge choice obeys the corresponding projective symmetry group transformations that combines a physical symmetry (here, translation or threefold rotation) and an appropriate gauge transformation \cite{Wen}. Consequently, no symmetries other than the time reversal  are broken in the physical ground state. 

For the system with cylindrical symmetry, Fig. \ref{edge}(a) and (b) illustrate the existence of chiral edge modes indicating a finite spectral Chern number of $\pm 2$. For $\Ju=\Ju'$, $\Jd=\Jd'$, and $J_5=0$, the $c$ and $d$ Majorana fermions decouple and have identical spectrum. It follows that the two gapless chiral Majorana edge states overlap as shown in Fig. \ref{edge} (a). Generically, the two chiral edge modes are separated, as shown in Fig. \ref{edge}(b). Moreover, the arguments in Refs. \cite{Baskaran:prl07, Wu:prb09, Yao:prl09} for spin-correlations apply here as well--the spin correlations vanish beyond nearest neighbors, establishing a CSL phase with a finite spectral Chern number and gapless chiral edge states on the boundary as a ground state of \eqref{eq:H_gen}. Because of the even spectral Chern number ($\pm 2$), we expect the vortex excitations are Abelian. Nonetheless, if the parameters in the Hamiltonian doubles the unit cell, a gapped non-Abelian chiral spin liquid can also be obtained \cite{Chua:10}.

\begin{figure}[bt]
\includegraphics[scale=0.5]{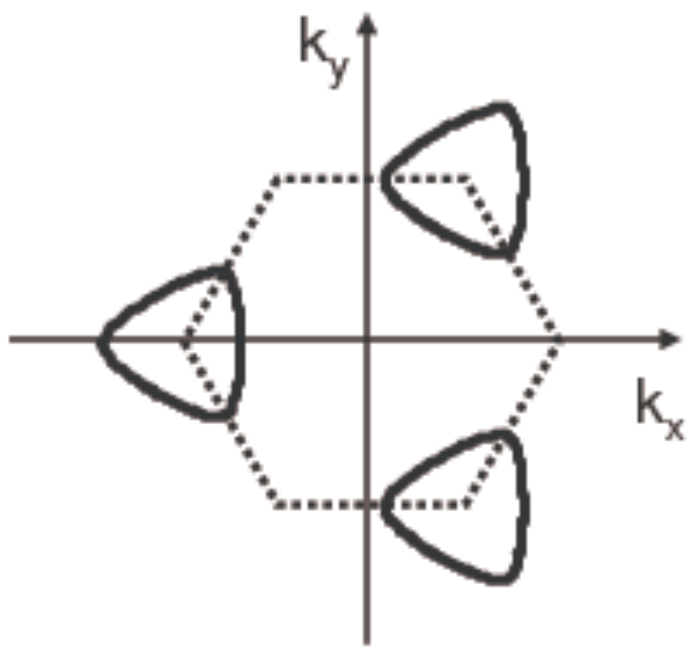}
\caption{The Fermi surface (solid line) for the flux configuration $\{\frac{\pi}{2},\frac{\pi}{2},\pi\}$ and  $\{\Ju,\Jd,\Ju',\Jd',J_5\}=\{1.0, 0.3, 0.8, 0.5,1.4\}$. The dashed hexagon is the Brillouin zone boundary.  Note that there is only one Fermi pocket for this set of parameters and the three pockets shown are related by a reciprocal lattice vector and are thus equivalent. } 
\label{fig3}
\end{figure}

{\it Gapless spin liquid with stable spin Fermi surface--} 
For arbitrary values of $\Ju,\Jd,\Ju',\Jd'$ in \eqref{eq:H_gen}, we found that the system is always gapped so long as we kept $J_5=0$, where $W_p=-1$ for all $p$ \cite{Chua:10}. By increasing $J_5$, we found that the system closes its gap at some critical values of $J_5$ and 
the ground state flux configuration changes when a critical point is crossed [see Fig.~\ref{fig4}(a)]. Note that the energy differences between different flux states are surprising small, as shown in Fig. \ref{fig4}(a). For a generic $J_5$ not at those critical points, the system stays gapped, which seems to make the search for a spin Fermi surface difficult for the Hamiltonian (1). Fortunately, while changing $J_5$ a particular flux configuration can be always favored by adding a pure flux term that does not destroy the exact solvability of the model. Specifically, to stabilize a flux configuration 
we add the following term to Eq.~\eqref{eq:H_gen},
\begin{equation}
\label{eq:H_FS}
{\cal H}_{\rm FS}=-\alpha\sum_{\hexagon} \hat W_{\hexagon} -\beta\sum_{\langle\triangle,\nabla\rangle}\hat W_{\triangle}\hat W_{\nabla}.
\end{equation}
The required value of $\alpha$ and $\beta$ to stabilize a desired flux pattern depends on $J_5$, but is typically rather small \cite{Tikhonov:prl10,Baskaran:09} as the energy differences between different flux states are small--see Fig. \ref{fig4}(a).  

For appropriate positive $\alpha$ and $\beta$, the ground state with the flux configuration $\{\frac{\pi}{2},\frac{\pi}{2},\pi\}$ (or  $\{-\frac{\pi}{2},-\frac{\pi}{2},\pi\}$) 
is favored. For this flux configuration, we obtain a finite Fermi surface as shown in Fig.\ref{fig3} (a) for $\Ju=1.0$, $\Jd=0.3$, $\Ju'=0.8$, $\Jd'=0.5$, and $J_5=1.4$. The dashed line of the hexagon is the Brillouin zone boundary. Note that there is only one Fermi pocket around one inequivalent corner for this set of parameters. The Fermi surface has a $C_3$ rotational symmetry around the zone corner that is expected from the $C_3$ symmetry of the model. Varying the parameters in Eq. (1) by a small amount will only change the shape and size of the Fermi surface but the Fermi surface itself is robust. 

An important question to ask about a Fermi surface is whether it is stable against {\it any} weak perturbations. In 2D, a generic Fermi surface can not be fully gapped by a weak ``density wave'' due to the lack of perfect nesting. However, a generic Fermi surface can still be fully gapped by a ``pairing'' term with momentum $\vec Q$ if the spectrum at $\vec k$ and $-\vec k+\vec Q$ are degenerate for any $\vec k$ and some fixed $\vec Q$.  In general, time reversal, $\pi$ rotation, or lattice inversion symmetry enforces degeneracy at $\vec k$ and $-\vec k+\vec Q$ and allows a putative Fermi surface to be gapped by an infinitesimal ``pairing'' term with momentum $\vec Q$. Note that a nonzero $\vec Q$ is encountered in the physical state but no gauge choice possesses one of those symmetries. If the ground state has no time reversal, $\pi$ rotation, or inversion symmetry, a putative Fermi surface will be {\it stable} against any weak perturbations. In the current model, the $\pi$ rotation and inversion symmetries are explicitly broken by $\Ju\neq \Jd$ or $\Ju'\neq \Jd'$; the time reversal symmetry is always spontaneously broken. It follows that the spin Fermi surface found in our model is stable against any weak perturbations. This is in contrast with the spin Fermi surface found in Refs. \cite{Yao:prl09,Baskaran:09,Tikhonov:prl10} that can be gapped by some sort of weak ``nesting'' or ``pairing'' terms. Lastly, this state has recently been shown to exhibit bond energy correlations that have $1/|{\bm r}|^3$ power law behaviour \cite{Lai} despite having ultra-short spin correlations.

{\it Concluding remarks--}We studied an exactly solvable model of spins on the kagome lattice with an odd number of electrons per unit cell (making it a ``true'' Mott insulator) that realizes a gapped chiral spin liquid with two gapless chiral Majorana edge modes and a gapless phase with a finite spin Fermi surface that is stable against any weak perturbations.  While in terms of the spin variables the model may seem highly anisotropic and therefore somewhat artificial, Kitaev-type models have been derived as effective low-energy theories from more ``realistic" Hamiltonians with strong spin-orbit coupling \cite{Jackeli:prl09} and can be readily engineered in optical lattices \cite{Roustekoski:prl09,Duan:prl03}.

We thank  M. Kargarian, D.-H. Lee, A. R\"uegg,  and J. Wen for helpful discussions. We are grateful to Steve Kivelson for importantly inspiring this work. This work is supported in part by NSF grant  DMR-0955778 (VC and GAF) at Austin and DOE grant DE-AC02- 05CH11231 (HY) at Berkeley.


\end{document}